\documentclass[twocolumn,secnumarabic,amssymb, nobibnotes, aps, prd]{revtex4-1}

\begin{document}

\def\TODAY{9 April 2010; 21 April 2010; 17 June 2010}

\title{\bf Kodama time: Geometrically preferred foliations of spherically symmetric spacetimes}

\author{Gabriel Abreu}%
\email{gabriel.abreu@msor.vuw.ac.nz}
\affiliation{School of Mathematics, Statistics, and Operations Research; Victoria University of Wellington, Wellington, New Zealand.}
\author{Matt Visser}%
\email{matt.visser@msor.vuw.ac.nz}
\affiliation{School of Mathematics, Statistics, and Operations Research; Victoria University of Wellington, Wellington, New Zealand.}

\date{\TODAY;  \LaTeX-ed \today}                                           
\begin{abstract}
In a general time-dependent (3+1)-dimensional spherically symmetric spacetime, the so-called Kodama vector is a naturally defined geometric quantity that is timelike outside the evolving horizon and so defines a preferred class of fiducial observers. However the Kodama vector does not \emph{by itself} define any preferred notion of time.  We first extract as much information as possible by invoking the ``warped product'' structure of spherically symmetric spacetime to study the Kodama vector, and the associated Kodama energy flux, in a coordinate independent manner. Using this formalism we  construct a general class of conservation laws, generalizing Kodama's energy flux. 

We then demonstrate that a  preferred time coordinate --- which we shall call Kodama time --- can be introduced by taking the additional step of applying the Clebsch decomposition theorem to the Kodama vector. We thus construct a  geometrically preferred coordinate system for any time-dependent spherically symmetric spacetime, and explore its properties. We study  the geometrically preferred fiducial observers, and demonstrate that it is possible to define and calculate a generalized notion of \emph{surface gravity} that is valid throughout the entire evolving spacetime.  Furthermore, by building and suitably normalizing set of radial null geodesics, we can show that this generalized surface gravity passes several consistency tests and has a physically appropriate static limit. 
\end{abstract}

\maketitle

\def\d{{\mathrm{d}}}
\newcommand{\scri}{\mathscr{I}}
\newcommand{\sun}{\ensuremath{\odot}}
\def\J{{\mathscr{J}}}
\def\sech{{\mathrm{sech}}}
\def\T{{\mathcal{T}}}
\def\B{{ {}^{\scriptscriptstyle B}\! }}
\def\F{{ {}^{\scriptscriptstyle F}\! }}
\section{Introduction}

Black holes are an iconic part of Einstein's general relativity. While we have a very detailed understanding of static and stationary black holes (the Schwarzschild, Reissner--Nordstr\"om, Kerr, and Kerr--Newman black holes), the situation with regard to evolving black holes, (evolving either due to accretion or Hawking radiation or both), is much more opaque. 
In particular, the rather limited number of currently known exact evolving solutions (Oppen\-heim\-er--Snyder collapse, the Vaidya solution) makes it much more difficult to fully describe an evolving black hole in any analytic detail. A  fundamental feature of the geometry of an evolving time-dependent spacetime is the lack of any (asymptotically timelike) Killing vector field, which seems to leave us without a preferred time coordinate with which to study the problem. 

In 1980 Kodama made significant progress in this regard when he constructed a \emph{geometrically natural} divergence-free \emph{preferred} vector field that is guaranteed to exist in any time-dependent spherically symmetric spacetime \cite{Kodama}.  This  so-called  ``Kodama vector''  defines a natural timelike direction in the region exterior to the black hole, and additionally induces an unexpected conserved current, but does not (in and of itself) define any naturally preferred time coordinate.  By considering the ``warped product'' form of the spacetime metric for any spherically symmetric geometry we are able to investigate the Kodama vector (and the associated Kodama energy flux) in a coordinate independent manner.  In particular we can use this formalism to easily generate a generalized Kodama flux.

By furthermore taking the extra step of invoking the Clebsch decomposition~\cite{lamb, deser, vorticity} on the (1+1) dimensional radial-temporal plane, we shall demonstrate that the Kodama vector field can indeed be used to construct a preferred time coordinate, and more importantly a preferred coordinate system.
The  absence of any (asymptotically timelike) Killing vector  in evolving spacetimes has made it difficult to achieve any consensus about best way to define such fundamental quantities as the surface gravity. Over the years, several different attempts have been made to extend the concept of surface gravity from static (and stationary) to time-dependent spacetimes. For instance, Hayward \cite{Hayward} uses the Kodama vector itself as a substitute for the Killing vector, since it certainly provides a preferred direction and it is parallel to the Killing vector in the static case (as well as at spatial infinity if one assumes the evolving spacetime is  asymptotically flat). Others (see \cite{Nielsen, Nielsen-Visser} and references therein) have appealed to the freedom of normalization of the null geodesics to ensure their definitions reduce to known results in the static case~\cite{putative}. 

The layout of the current article is as follows: In section \ref{S:warped} we briefly summarize key properties of ``warped product'' spacetimes. 
In section \ref{S:miracle} we present a quick review of the Kodama vector and Kodama's unexpected conservation law, and then significantly generalize Kodama's energy flux in section \ref{S:generalized}. Next, in section \ref{S:clebsch}, a Clebsch decomposition of the Kodama vector is made --- in order to build a natural geometrically preferred coordinate system for any spherically symmetric time-dependent spacetime.  In sections \ref{S:Riemann} and \ref{S:Einstein} we explore the Riemann and Einstein tensors in this geometrically preferred coordinate system, being careful to connect the discussion back to the general ``warped product'' formalism of section \ref{S:warped}. Furthermore, in section \ref{S:conservation} we review Kodama's conservation law in these preferred coordinates. In section \ref{S:brown-york} we calculate the Brown--York quasi-local mass, and in section and \ref{S:surface-gravity} present our extended definition of surface gravity. Section \ref{S:evolving-horizon} deals with the naturally induced notions of apparent and trapping horizon. Lastly, we add a brief discussion.

\section{Warped product spacetimes}\label{S:warped}

Any (possibly time-dependent) spherically symmetric metric can be written in the form
\begin{eqnarray}
\d s^2 &=& g_{ab} \; \d x^a \d x^b
\nonumber
\\
&=& \B g_{ij}(x) \; \d x^i\d x^j + r(x)^2 \, \left\{ \d{\theta}^2 + \sin^2{\theta} \;\d{\phi}^2 \right\}
\nonumber\\
&=&  \B g_{ij}(x) \; \d x^i\d x^j + r(x)^2 \; \F g_{\alpha\beta} \; \d x^\alpha \d x^\beta.
\end{eqnarray}
Here the two coordinates $x^i$ run over the radial-temporal plane, while the two coordinates $x^\alpha$ ($\theta$ and $\phi$) run over the surfaces of spherical symmetry. 
The discussion can be generalized to $(d+1)$ dimensions with $d-1$ dimensional spherical symmetry, but for now we are just working in (3+1) dimensions.  (For higher-dimensional generalizations in a Gauss--Bonnet context see~\cite{hideki1, hideki2}.) Independent of the total dimensionality,  $g_{ij}$ is a (1+1) dimensional Lorentzian metric. 

Geometrically this is called a ``warped product'' manifold, with the radial-temporal plane being referred to as the ``base space'', the surfaces of spherical symmetry being referred to as the ``fibres'', and the function $r(x)$ which depends only on the base space coordinates being referred to as the ``warp factor''. 
It is a standard computation to show that (up to the usual permutation symmetries for the indices) the only non-zero components of the Riemann tensor (in any warped product spacetime) are
\begin{eqnarray}
R_{ijkl} &=& \B R_{ijkl};
\\
R_{i\alpha j\beta} &=&  - \,  r \; \{ \nabla_i \nabla_j r \} \; \F g_{\alpha\beta};
\\
R_{\alpha\beta\mu\nu} &=&  r^2 \left\{  \F R_{\alpha\beta\mu\nu} 
 - {|\nabla r|^2} \left( \F g_{\alpha\mu} \F g_{\beta\nu} -  \F g_{\alpha\nu} \F g_{\beta\mu} \right) \right\}\!\!. \quad
\end{eqnarray}
An abstract computation along these lines can be found in O'Neill~\cite[page 210]{ONeill} while more explicit computations can be found in~\cite{Carot,  Agaoka, Senturk}. Note that the covariant derivatives appearing above are covariant derivatives in the base space.  But  because $r(x)$ depends only on the (1+1) dimensional base space coordinates, and because of the specific form of the warped product metric, these derivatives can be ``bootstrapped'' to  covariant derivatives in the total warped product spacetime.

In the specific situation we are interested in the base space is two dimensional, so in terms of the Ricci scalar of the radial-temporal plane we have the specific simplification
\begin{equation}
 \B R_{ijkl} =   {\B R\over2}   \; \left(    \B g_{ik} \B g_{jl} -  \B g_{il} \B g_{jk} \right).
\end{equation}
Furthermore the fibre is a constant curvature sphere of radius unity, so
\begin{equation}
 \F R_{\alpha\beta\mu\nu}
= \left( \F g_{\alpha\mu} \F g_{\beta\nu} -  \F g_{\alpha\nu} \F g_{\beta\mu} \right). 
\end{equation}
Thus we now have on purely geometrical grounds
\begin{eqnarray}
R_{ijkl} &=& { \B R\over2}  \; \left(    \B g_{ik} \B g_{jl} -  \B g_{il} \B g_{jk} \right);
\\
R_{i\alpha j\beta} &=&  - \, r \; \{\nabla_i \nabla_j r\} \;  \F g_{\alpha\beta};
\\
R_{\alpha\beta\mu\nu} &=&  r^2 \; \left\{ 1 -  {|\nabla r|^2} \right\} \; \left( \F g_{\alpha\mu} \F g_{\beta\nu} -  \F g_{\alpha\nu} \F g_{\beta\mu} \right).
\end{eqnarray}
It is also common in spherical symmetry to define the Hawking--Israel/ Hernandez--Misner/ Misner--Sharp quasi-local mass~\cite{hernandez-misner, misner-sharp} by
\begin{equation}
1- {2m\over r} = |\nabla r|^2
\end{equation}
where both $m(x^i)$ and $r(x^i)$ are scalar functions on the base space.
We now have
\begin{eqnarray}
R_{ijkl} &=& {\B R\over2}  \; \left(    \B g_{ik} \B g_{jl} -  \B g_{il} \B g_{jk} \right);
\\
R_{i\alpha j\beta} &=&  -\,  r \; \{\nabla_i \nabla_j r\} \;  \F g_{\alpha\beta};
\\
R_{\alpha\beta\mu\nu} &=&   2 m \; r\;  \; \left( \F g_{\alpha\mu} \F g_{\beta\nu} -  \F g_{\alpha\nu} \F g_{\beta\mu} \right).
\end{eqnarray}
It is often useful to go to an orthonormal basis, in which case
\begin{eqnarray}
R_{\hat i\hat j\hat k\hat l} &=& {\B R\over2}  \; 
\left(    \delta_{\hat i\hat k} \; \delta_{\hat j\hat l} -  \delta_{\hat i\hat l} \; \delta_{\hat j\hat k} \right);
\\
R_{\hat i\hat\alpha\hat j\hat \beta} &=&  -\,  {\{\nabla_{\hat i} \nabla_{\hat j} r\}\over r}  \;  \delta_{\hat\alpha\hat\beta};
\\
R_{\hat\alpha\hat\beta\hat\mu\hat\nu} &=&   {2 m\over r^3}\;  \; 
\left( \delta_{\hat\alpha\hat\mu} \; \delta_{\hat\beta\hat\nu} -  \delta_{\hat\alpha\hat\nu} \; \delta_{\hat\beta\hat\mu} \right).
\end{eqnarray}
For the Ricci tensor we have
\begin{eqnarray}
R_{ij} &=&  {\B R\over2}  \;  \B g_{ij}  - 2  {\{\nabla_i \nabla_j r\} \over r};
 \\
R_{i\alpha} &=& 0;
\\
R_{\alpha\beta} &=& \left\{  {2m\over r}  - { r \nabla^2} r \right\}  \F g_{\alpha\beta };
\end{eqnarray}
and for the Ricci scalar
\begin{equation}
R =  \B R  - 4 { \nabla^2 r\over r} + {4m\over r^3},
\end{equation}
where the Laplacians above are in the (1+1) dimensional sense on the base space.
The Einstein tensor takes the form
\begin{eqnarray}
G_{ij} &=& -   {2\{\nabla_i \nabla_j r\} \over r} + \left\{  { 2 \nabla^2 r \over r} - {2m\over r^3} \right\}\;  \B g_{ij};
  \\
G_{i\alpha} &=& 0;
\\
G_{\alpha\beta} &=& \left\{ -  {\B R \; r^2 \over2}  + {  r \nabla^2 r } \right\} \;  \F g_{\alpha\beta }.
\end{eqnarray}
These are all purely geometrical statements --- while one has chosen coordinates $x^a = (x^i; x^\alpha)$ to make the warped product structure manifest, these results are completely independent of one's choice of coordinates $x^i$ on the radial-temporal plane (the base space), and for that matter are completely independent of one's choice of coordinates $x^\alpha$ on the spherically symmetric fibres.

\section{The Kodama miracle}\label{S:miracle}
It is well known that in a  time-dependent spacetime, there is no (asymptotically timelike) Killing vector to define a preferred time coordinate. The calculation of important quantities such as the four-acceleration and the surface gravity become much more ambiguous. Additionally,  there is no general consensus  on the ``best'' form of the metric, nor on the ``best'' choice of the coordinate system. 

An interesting insight on this problem is  given by Kodama \cite{Kodama}, who proved the existence of a divergence-free vector field for any time-dependent spherically symmetric metric.
The Kodama vector, $k^a$, lies in the (1+1) dimensional radial-temporal plane, so that $k^a = (k^i; \,0,0)$. More precisely
\begin{equation}
k^a = \epsilon_\perp^{ab} \; \nabla_b r.
\end{equation}
where the tensor $\epsilon_\perp^{ab}$ is the (1+1) dimensional Levi--Civita tensor in the radial-temporal plane, denoted $\epsilon_\perp^{ij}$, canonically embedded into (3+1) dimensions according to the prescription
\begin{equation}
\epsilon_\perp^{ab} = \left[\begin{array}{c|c} \vphantom{\Big{|}} \epsilon_\perp^{ij} & 0 \\ \hline 0 & 0 \end{array}\right].
\end{equation}
It is straightforward to check that $ k^a \; \nabla_a r = 0$. Furthermore, if we define a positive semi-definite norm by $||k||^2 = |g(k,k)| = |g^{-1}(k^\flat,k^\flat)|$, then  $||k|| = ||\nabla r||$. (We shall use the superscripted symbol $\flat$ to denote the process of turning a vector into a covector by ``lowering the index'', and use the superscripted symbol $\sharp$ to denote the inverse process of turning a covector into a vector by ``raising the index''.) By appropriate choice of orientation on the radial-temporal plane one can choose $k^a$ to be (asymptotically) future pointing. It can also be defined by the more  abstract statement
\begin{equation}
k = ( *_2  \; \d r)^\sharp,
\end{equation}
where by this one means ``calculate the one-form $\d r$, apply the (1+1) dimensional radial-temporal Hodge star operation, and use the metric to convert the resulting one-form to a contravariant vector''. 
At this point it is necessary to emphasise, as originally pointed out by Kodama himself, that the Kodama vector does \emph{not} in general reduce to the Killing vector in a static spacetime; all that one can say in general is that in static spacetimes it is \emph{parallel} to the Killing vector. In regions where the Kodama vector is timelike (and we shall [informally at this stage] refer to this as the black hole exterior region, \emph{i.e.}, the domain of outer communication) the Kodama vector defines a preferred class of fiducial observers (FIDOs)~\cite{membrane} specified by the unit timelike 4-vector
\begin{equation}
V = { k\over ||k||}.
\end{equation}
Although the Kodama vector provides a preferred ``time direction'',  and simplifies the evolution equations of a dynamical spherically symmetric system \cite{Racz1, Racz2}, it does not at this stage define a preferred ``time coordinate''. We shall subsequently use the Kodama vector plus the the Clebsch decomposition theorem to construct a geometrically natural preferred time coordinate. 

To prove that the Kodama vector is divergence free the best strategy (with hindsight) is to consider the quantity
\begin{eqnarray}
\nabla_a( \epsilon_\perp^{ab}/r^2) &=& {1\over\sqrt{-g_4}} \partial_a \left( \sqrt{-g_4}  \; \epsilon_\perp^{ab}/r^2 \right) 
\nonumber\\
&=&  
 {1\over\ r^2 \,\sqrt{-g_2}  } \partial_a \left( \sqrt{-g_2}  \;  \left[\begin{array}{c|c} \vphantom{\Big{|}} \epsilon_\perp^{ij} & 0 \\ \hline 0 & 0 \end{array}\right] \right) 
 \nonumber\\
&=&  
\left(  {1\over r^2 \, \sqrt{-g_2}  } \partial_i \left[ \sqrt{-g_2}  \; \epsilon_\perp^{ij}   \right] ; 0, 0 \right)
 \nonumber\\
 &=&  
{1\over r^2} \left( \B\, \nabla_i \epsilon_\perp^{ij} ; 0, 0 \right)
 \nonumber\\
 &=&
 0.
\end{eqnarray}
Note that the last covariant derivative is a base space covariant derivative which vanishes since the (1+1) dimensional Levi-Civiata tensor is covariantly constant with respect to the (1+1) dimensional covariant derivative. But this (3+1) dimensional result, $\nabla_a( \epsilon_\perp^{ab}/r^2)=0$ can easily be rearranged to give
\begin{equation}
k^a =  {r\over2} \; \nabla_b \epsilon_\perp^{ab},
\end{equation}
which now implies
\begin{equation}
\nabla_a k^a = 0,
\end{equation}
so the Kodama vector itself is conserved. 
In addition, Kodama also proved that in time-dependent spherically symmetric spacetimes there is another (somewhat unexpected) conserved current. In terms of the Einstein tensor and the Kodama vector we have  $J^a = G^{ab} \; k_b$, with:
\begin{equation}
\nabla_a J^a=\nabla_a(G^{ab}\;k_b)=0.
\end{equation}
This is a purely geometrical statement, ultimately due to the warped product form of the metric --- it is \emph{not} related to the Bianchi identities. 
Let us specifically compute
\begin{equation}
J^a = G^{ab} \; k_b = ( G^{ij} k_j; \; 0,0).
\end{equation}
Now working on the radial-temporal base space we have
\begin{equation}
 G_{ij} k^j = - 2  {\nabla_i \nabla_j r\over r}  \epsilon_\perp^{jk} \nabla_k r + \left\{ {2 \nabla^2 r\over r} - {2m\over r^3} \right\}\;  k^i.
\end{equation}
But using the fact that in (1+1) dimensions 
\begin{equation}
 \nabla_{[i} r  \; \epsilon_{\perp jk]} = 0,
\end{equation}
we have
\begin{eqnarray}
 {\nabla_i \nabla_j r}  \; \epsilon_\perp^{jk} \; \nabla_k r  &=&  {\nabla^j \nabla_i r}  \; \epsilon_{\perp jk} \; \nabla^k r 
\nonumber
 \\
&=&   \nabla^j  \{ \nabla_i r  \; \epsilon_{\perp jk} \} \; \nabla^k r  
\nonumber
\\
&=&   \nabla^j  \{ -\nabla_j r  \; \epsilon_{\perp ki}  - \nabla_k r  \; \epsilon_{\perp ij} \} \; \nabla^k r  \qquad
\nonumber\\
&=&+  {\nabla^2 r}\; k_i  - {1\over2 }  \epsilon_{\perp ij}  \nabla^j \{ |\nabla r|^2\}.
\end{eqnarray}
Combining these results
\begin{equation}
 G_{ij} k^j = + {1\over r}  \epsilon_{\perp ij}  \nabla^j \{ |\nabla r|^2\} - {2m\over r^3} \;  k^i.
\end{equation}
But in view of the definition of the  Hawking--Israel/ Hernandez--Misner/ Misner--Sharp quasi-local mass~\cite{hernandez-misner, misner-sharp} we then have
\begin{equation}
 G_{ij} \, k^j =  -{1\over r}  \epsilon_{\perp ij}  \nabla^j \{ 2m/r \} - {2m\over r^3} \;  k^i.
\end{equation}
That is
\begin{equation}
 G_{ij} \; k^j =  - {2\over r^2}  \epsilon_{\perp ij}  \nabla^j m.
\end{equation}
This computation has been performed using the (1+1) dimensional covariant derivative in the radial-temporal base space, but at this stage we can safely use the symmetries of the situation to lift this equality to the full spacetime
\begin{equation}
J^a =  G^{ab} k_b =  - {2\over r^2} \;  \epsilon_{\perp}^{ab} \;  \nabla_b m.
\end{equation}
This is a purely geometrical statement --- fundamentally connected with the warped product nature of the spacetime --- that underlies the unexpected conservation of the Kodama current. In view of the fact that we have already proven $\nabla_a (\epsilon_\perp^{ab}/r^2)=0$ we finally see
\begin{equation}
\nabla_a J^a =  - {2\over r^2} \;  \epsilon_{\perp}^{ab} \;  \nabla_a \nabla_b m  = 0.
\end{equation}
Thus conservation of the Kodama flux is is a subtle result deeply connected with the warped product nature of the spacetime. We have presented this derivation in some detail because it is now possible to rapidly generalize the result in a significant manner.

\section{Generalized Kodama flux}\label{S:generalized}

Consider an arbitrary function $\Phi(m,r)$ of the two quantities $m(x^i)$ and $r(x^i)$. Now construct the current
\begin{equation}
J_\Phi^a =  \left\{ \partial_m \Phi(m,r) \; G^{ab} -2 \,\partial_r\Phi(m,r) \; g^{ab} \right\} k_a.
\end{equation}
This current is conserved in any spherically symmetric spacetime. To prove this note that by the definition of the Kodama vector and the geometrical identity proved above we have
\begin{eqnarray}
J_\Phi^a &=&  -2 {\epsilon_\perp^{ab}\over r^2}  \left\{ \partial_m \Phi(m,r) \;  \nabla_b m+ \partial_r\Phi(m,r)  \nabla_b r \right\}
\nonumber\\
&=&   -2 {\epsilon_\perp^{ab}\over r^2} \nabla_b \Phi(m,r).
\end{eqnarray}
Conservation of this 4-vector is then obvious from the last expression.  Note in particular that by the above argument any flux  of the form
\begin{equation}
J_{12}^a =  \left\{ f_1(m) \; G^{ab} + f_2(r) \; g^{ab} \right\} k_a
\end{equation}
[for arbitrary $f_1(m)$ and $f_2(r)$]
will automatically be conserved. 

Formally, there is an even more general conserved current one can write down: For any arbitrary scalar function $\Psi(x^i)$ defined on the radial-temporal base space the quantity 
\begin{equation}
\label{ConservSet}
J_\Psi^a =   {\epsilon_\perp^{ab} \over r^2} \; \nabla_b \Psi,
\end{equation}
is conserved.  Though this result is more general, it is somewhat less geometrical, and does not have the same flavour as the above. If (and only if) the functions $r(x^i)$ and $m(x^i)$ are functionally independent (so that one can use $m$ and $r$ as coordinates on the radial-temporal base space) then these two notions ($J_\Phi$ and $J_\Psi$) can be made to coincide.  In particular, $\Psi \to - 2 m(r,t)$ gives us Kodama's conserved flux $J^a$, while $\Psi \to {1\over 3} r^3$ is just the statement that the Kodama vector itself is conserved, $\nabla_a k^a =  0$.  (For related comments in a higher-dimensional Gauss--Bonnet context see~\cite{hideki1, hideki2}.)

\section{Kodama time}\label{S:clebsch}
The Kodama vector has been used before in several aspects of the time-dependent gravitational collapse problem. However it has not been used to obtain a preferred time coordinate, nor a preferred coordinate system for the metric of a dynamic spacetime.
Fortunately, in (1+1) dimensions it is possible to use the lesser known but classic Clebsch decomposition theorem, a result complementary to the more usual Helmholtz decomposition theorem,  (see for instance~\cite{lamb, deser, vorticity}) to assert that there are two unique scalars $\alpha$ and $\beta$ such that the Kodama covector $k^\flat$ takes the form
\begin{equation}
k^\flat = \alpha \; \d\beta.
\end{equation}
Now in the ``normal'' exterior region where $\d r$ is spacelike,  (\emph{i.e.}, in the domain of outer communication), the Kodama vector and covector are both timelike, so in this region the one-form $\d\beta$ is guaranteed to be timelike. This \emph{very strongly suggests} that $\beta$ should be adopted as a preferred ``time coordinate''. In fact, relabeling  the scalar $\beta$ as a coordinate $\beta\to t$, and relabeling the integrating factor as $\alpha\to F$, we write
\begin{equation}
k^\flat = F \; \d t.
\end{equation}
This naturally induces a geometrically preferred time coordinate $t$ --- which we shall refer to as the \emph{Kodama time}. Using this time coordinate is at least as natural as using $r$ for the radial coordinate.
(This key step, though mathematically elementary,  goes well beyond anything in Kodama's original paper~\cite{Kodama}, or the various papers that have subsequently sought to use Kodama's formalism.) As we shall soon see, this choice of Kodama time coordinate is the unique choice that makes integral curves of the vector $\partial_t$ coincide with integral curves of the Kodama vector. That is
\begin{equation}
k \propto \partial_t.
\end{equation}
Ultimately, adopting these coordinates (no matter how natural they appear) is of course a \emph{choice}, and will be ``justified'' only insofar as they turn out to be useful. 

Adopting these ($t$, $r$) coordinates as preferred coordinates on the radial-temporal plane, and without any loss of generality, the metric can be written as
\begin{eqnarray}
\d s^2 =  g_{tt}(r,t) \, \d t^2  + 2 g_{tr}(r,t) \d r \,\d t + g_{rr}(r,t) \,\d r^2
\nonumber\\
+ r^2 \, \left\{ \d{\theta}^2 + \sin^2{\theta} \;\d{\phi}^2 \right\}.
\end{eqnarray}
However, since the Kodama vector $k$ is orthogonal to $\d r$, then also $\d t$ is orthogonal to $\d r$, and so the cross term in the metric is zero. 
Thus in these preferred coordinates the metric is diagonal,
\begin{equation}
\d s^2 =  g_{tt}(r,t) \, \d t^2 + g_{rr}(r,t) \,\d r^2+ r^2 \, \left\{ \d{\theta}^2 + \sin^2{\theta} \;\d{\phi}^2 \right\}.
\end{equation}
This relatively long argument has ultimately led us back to one of the simplest, and arguably most obvious, forms of the metric --- a simple diagonal metric in Schwarzschild curvature coordinates. (Of course now we can argue that we have a \emph{geometrically natural reason} for adopting this particular set of coordinates.) 

When using the Schwarzschild radial coordinate $r$  it is natural to choose the parameterization
\begin{equation}
 g_{rr}(r,t) =  \left(1 - {2m(r,t)\over r}\right)^{-1}
 \end{equation}
for the radial-radial part of the metric tensor. Doing so will automatically give the quantity $m(r,t)$ a natural interpretation in terms of the Hawking--Israel/ Hernandez--Misner/ Misner-Sharp  quasi-local mass~\cite{hernandez-misner, misner-sharp}.
Since the radial-temporal plane by definition has Lorentzian signature, this choice then guarantees that it is possible to write the temporal-temporal  component of the metric tensor in the form
\begin{equation}
g_{tt}(r,t) =  - e^{-2\Phi(r,t)} \left(1 - {2m(r,t)\over r}\right).
\end{equation}
We finally have the (quite standard) result
\begin{eqnarray}
\label{met}
\d s^2 &=& 
- e^{-2\Phi(r,t)} \left(1 - {2m(r,t)\over r}\right) \, \d t^2 + {\d r^2 \over 1 - 2m(r,t)/r} 
\nonumber\\
&&
\qquad\qquad
+  r^2 \, \left\{ \d{\theta}^2 + \sin^2{\theta} \;\d{\phi}^2 \right\},
\end{eqnarray}
where in addition we know 
\begin{equation}
\nabla_a r = (\d r)_a=  (0,1;\, 0,0),
\end{equation} 
and the equivalent contravariant  result
\begin{equation}
(\d r)^a = \left(0, 1-{2m(r,t)\over r}; \, 0,0\right).
\end{equation}
Furthermore, the components of the Kodama vector and covector in these coordinates are
\begin{eqnarray}
k^a &=& e^{\Phi(r,t)} (1,0; \,0,0);  \qquad 
\nonumber\\
 k_a &=&  - e^{-\Phi(r,t)}  \left(1 - {2m(r,t)\over r}\right) (1,0; \,0,0) .
\end{eqnarray}
As previously mentioned, the squared norm of the Kodama vector is equal to that of $\nabla r$:
\begin{equation}
||k||^2 = ||\nabla r||^2 = \left|1 - {2m(r,t)\over r}\right|.
\end{equation}
In these coordinates it is useful to define the time translation vector, $\T$, \emph{which is not a Killing vector unless the geometry happens to be static}, as
\begin{eqnarray}
\T = \partial_t;  \qquad  \T^a = (1,0; \,0,0); \qquad 
\nonumber\\
 \T_a = e^{-2\Phi(r,t)}  \left(1 - {2m(r,t)\over r}\right) (1,0; \, 0,0).
\end{eqnarray}
The squared norm of $\T$ is equal to the absolute value of the temporal-temporal component of the metric, $|g_{tt}|$, and proportional to the squared norm of the Kodama vector:
\begin{equation}
 ||\T||^2 = |g_{tt}| =    e^{-2\Phi(r,t)} \left|1 - {2m(r,t)\over r}\right| =  e^{-2\Phi(r,t)}\;||k||^2.
\end{equation}
That is,
\begin{equation}
\label{expFactor}
 e^{-2\Phi(r,t)}  = { ||\T||^2\over ||k||^2}; \qquad   e^{-\Phi(r,t)}  = { ||\T||\over ||k||}.
\end{equation}
In the static  situation it is the time translation vector $\T$ that will reduce to the (asymptotic) timelike Killing vector: $\T \to K \neq k$. Because the normalizations of $\T$ and $k$ differ, then if  the Kodama vector is used simply as a substitute for the Killing vector when attempting to calculate quantities such as the surface gravity~\cite{Hayward}, one is likely to encounter normalization issues when taking the static limit. 

Specifically, to obtain a finite value of the four-acceleration near a possible horizon, as measured by an observer at infinity, it is necessary to multiply by a suitably defined normalizing factor~\cite{carroll, wald}. In the static case this normalizing  factor is just $|{g}_{tt}|$, and coincides with the squared norm of the Killing vector. However, in the time dependent case, not only does the geometry  not possess a Killing vector, but also the squared norm of the Kodama vector does not coincide with $|g_{tt}|$. This leaves us with a somewhat ambiguous situation with respect to the normalizing factor and the surface gravity of a time-dependent metric tensor, and means that we will have to exercise some care in defining the surface gravity of a time-dependent geometry.

\section{Riemann tensor}\label{S:Riemann}

It is now a standard exercise to calculate the various components of the Riemann tensor (for instance, by using {\sf Maple}).  
We note that the Riemann tensor is considerably less fearsome than one might suppose. Only one component is in any sense ``difficult''. Temporarily suppressing the $(r,t)$ arguments for conciseness, and working in an orthonormal basis we have:
\begin{eqnarray}
R_{\hat t\hat r\hat t\hat r} &=&  - {2m\over r^3} - {m''\over r} +{2m'\over r^2} 
 \qquad
\nonumber\\
&&
+ \left( 1- {2m\over r} \right) \left[ -\Phi'' + (\Phi')^2 \right] + 3(m/r)' \, \Phi' 
\nonumber\\
&& - {e^{\Phi}\over r} \; \partial_t \left[ {\dot m \; e^\Phi \over  \left(1-{2m\over r}\right)^2}\right]. 
\end{eqnarray}
In view of the warped product formalism, we know that this rather messy quantity has a drect and simple physical/ mathematical interpretation: As may be verified by direct computation it is simply $\B R/2$, one half the Ricci scalar of the (1+1) dimensional radial-temporal plane. 

The remaining components are much simpler:
\begin{equation}
R_{\hat t\hat \theta\hat t\hat \theta } =  R_{\hat t\hat \phi\hat t\hat \phi } =  { m - r m'\over r^3 } - \left( 1- {2m\over r} \right) {\Phi'\over r};
\end{equation}
\begin{equation}
R_{\hat t\hat \theta\hat r\hat \theta } =  R_{\hat t\hat \phi\hat r\hat \phi } =  {\dot m e^\Phi\over r^2(1-2m/r) };
\end{equation}
\begin{equation}
R_{\hat r\hat \theta\hat r\hat \theta } =  R_{\hat r\hat \phi\hat r\hat \phi } =  -{ m - r m'\over r^3 }.
\end{equation}
These three quantities are easily seen to be proportional to $\nabla_i\nabla_j r $. (In fact they equal $- \{\nabla_{\hat i}\nabla_{\hat j} r\}/r$.)
Finally
\begin{equation}
R_{\hat \theta\hat \phi\hat \theta\hat\phi } =  { 2m \over r^3 }.
\end{equation}
Note that the particularly simple formula for $R_{\hat \theta\hat \phi\hat \theta\hat\phi }$ underlies the identification of $m(r,t)$ as the Hernandez--Misner quasi-local mass~\cite{hernandez-misner}.

\section{Einstein tensor}\label{S:Einstein}

For  the Einstein tensor, working in an orthonormal basis, the  single most important result is
\begin{equation}
G_{\hat t\hat t}= {2\,m'(r,t)\over r^2}.
\end{equation}
Here the primes denote differentiation with respect to $r$, and the dots with respect to $t$. 
This result is utterly standard, with the only novelty being that this formula for $G_{\hat t\hat t}$ continues to hold in the time dependent case (subject of course to the coordinate choices made above).  This result for $G_{\hat t\hat t}$ is intimately related to the physical interpretation of $m(r,t)$ as the Hawking--Israel quasi-local mass.  

A second important result is more subtle:
\begin{equation}
G_{\hat t\hat r} = {2\, \dot m(r,t) \, e^{\Phi(r,t)} \over r^2\,\left(1-{2m(r,t)\over r}\right)}.
\end{equation}
We shall soon see that this formula for $G_{\hat t\hat r}$ is central to the coordinate-based verification of Kodama's unexpected conservation law, and that it is intimately related to the Brown--York quasi-local mass~\cite{brown-york}.

For completeness we indicate
\begin{equation}
G_{\hat r\hat r}= - {2\,m'(r,t)\over r^2} - {2\Phi'(r,t) \,\left(1-{2m(r,t)\over r}\right)\over r}.
\end{equation}
This result is quite standard, (see for instance equation (2.65) of~\cite{traversable}), with the only novelty being that this formula for $G_{\hat r\hat r}$ continues to hold in the time dependent case. 
This now implies the useful result
\begin{equation}
G_{\hat t\hat t} + G_{\hat r\hat r}=  - {2\Phi'(r,t) \,\left(1-{2m(r,t)\over r}\right)\over r}.
\end{equation}

Finally, now suppressing the $(r,t)$ arguments for conciseness, we have
\begin{eqnarray}
G_{\hat\theta\hat \theta} = G_{\hat\phi\hat \phi} &=& \left(1-{2m\over r} \right) \left[ - \Phi'' + \Phi' \left( \Phi' - {1\over r} \right) \right] 
 \qquad
\\
&&
- 3 \Phi' \left( {m\over r^2} - {m'\over r} \right) - {m''\over r} \qquad
\nonumber\\
&&
- {e^{2\Phi}\over \left(1-{2m\over r}\right)^2 } \left[ {\ddot m\over r} + {4(\dot m)^2\over r \left(1-{2m\over r}\right)} - {\dot \Phi \, \dot m\over r} \right].
\nonumber
\end{eqnarray}
The first two lines here are again quite standard, and appear also in static situations. (See for instance equation (2.66) of~\cite{traversable}). All the time derivatives have been isolated in the third line.  
With a little more work this can be somewhat tidied up as follows
\begin{eqnarray}
G_{\hat\theta\hat \theta} &=& G_{\hat\phi\hat \phi} = 
 - {m''\over r} 
  \qquad
\nonumber\\
&&
 - {e^\Phi\over r \sqrt{1-2m/r}}\;  \partial_r \left[ r  \left(1-{2m\over r}\right)^{3/2} \; e^{-\Phi}\; \Phi' \right]
 \qquad
\nonumber\\
&&
- {e^{\Phi}\over r} \; \partial_t \left[ {\dot m \; e^\Phi \over  \left(1-{2m\over r}\right)^2}\right].
\end{eqnarray}
This rather complicated expression can be verified to equal $-\B R/2 + \nabla^2 r /r$, the result we expect based on the warped product formalism.
We note that  the time derivative contributions to the Einstein tensor are quite isolated, and in this geometrically preferred coordinate system occur only in the $G_{\hat\theta\hat \theta} = G_{\hat\phi\hat \phi}$ and $G_{\hat t\hat r}$ components.  This ultimately one of the key reasons we will find the Kodama time to be so useful.

\section{Co-ordinate based version of Kodama's conservation law}
\label{S:conservation}

Kodama's conservation law can now be studied in more explicit coordinate-based detail. First, based only on spherical symmetry and the definition of the Kodama vector, the unexpected conserved current $J^a$ takes the form
\begin{equation}
J^a = g^{ab} \; G_{bc} \; k^c = \left\{ -\hat k^a \hat k^b + (\widehat{\d r})^a   (\widehat{\d r})^b  \right\} G_{bc} \; \hat k^c \; ||k||,
\end{equation}
whence, since by construction $||k|| = ||\d r||$, we see
\begin{equation}
J^a = - G_{\hat t\hat t} \; k^a + G_{\hat t\hat r} \; (\d r)^a.
\end{equation}
But we have already explicitly calculated the quantities $G_{\hat t\hat t}$, $G_{\hat t\hat r}$, $k^a$ and $(\d r)^a$. We obtain
\begin{equation}
J^a = 2 \left( - {e^{\Phi(r,t)} m'(r,t)\over r^2}, {e^{\Phi(r,t)} \dot m(r,t)\over r^2}; \, 0 , 0\right).
\end{equation}
This vector is is now \emph{obviously} conserved since the 4-divergence is simply
\begin{eqnarray}
\nabla_a J^a &=&  {1\over\sqrt{-g_4}} \; \partial_a [ \sqrt{-g_4}\; J^a]
\nonumber\\
&=& {2\over e^{-\Phi(r,t)} \, r^2} \; \partial_a \left[ \left( - m'(r,t), \dot m(r,t); \, 0 , 0\right)^a \right]
\nonumber\\
&=&   {2\over e^{-\Phi(r,t)} \, r^2} \; [-\dot m'(r,t) + \dot m'(r,t)] 
\nonumber\\
&=& \vphantom{\Big|} 0.
 \end{eqnarray}
Equivalently we note that from this coordinate-based calculation we explicitly recover
\begin{equation}
\label{E:extra}
J^a =  G^{ab} \, k_b =  -2 \; {\epsilon_\perp^{ab} \; \nabla_b m\over r^2}.
\end{equation}
This relation is somewhat miraculous in the present coordinate based calculation, and as we have seen has a deeper justification in terms of the warped product form of the spacetime geometry. 

\section{Brown--York quasi-local mass}\label{S:brown-york}
With the coordinate system developed above, the notion of quasi-local internal energy arises naturally as the Brown--York quasi-local mass \cite{brown-york}. To prove this, first let us take some imaginary spherical surface surface $r=r_0$, and hold $r_0$ fixed in time. Then the total energy inside this spherical surface depends on the net flux across the surface. To calculate the net flux we need the $G_{tr}$ component of the Einstein tensor, in an orthonormal basis. That is
\begin{equation}
G_{\hat t\hat r}(r,t) = {2 \dot m(r,t)\over r^2(1-2m(r,t)/r)}\,e^{\Phi(r,t)},
\end{equation}
whence, via the Einstein equations $8\pi\,G_{ab}=T_{ab}$, we have the flux density
\begin{equation}
f(r,t)=   T_{\hat t \hat r}(r,t) = {1\over 4\,\pi}\,{ \dot m(r,t)\over r^2(1-2m(r,t)/r)}\,e^{\Phi(r,t)}.
\end{equation}
Now the total net flux across the imaginary surface at $r=r_0$, in an amount of proper time $\tau$, is
\begin{eqnarray}
\hbox{(net flux)}_{t_\mathrm{initial}}^{t_\mathrm{final}} &=&\int f(r_0,t) \times \hbox{(area)} \times \d\tau\nonumber\\
&=&\int f(r_0,t) \times(4\pi\,r_0^2) 
 \qquad
\nonumber\\
&& \times \left[e^{-\Phi(r_0,t)}\sqrt{1-\frac{2\,m(r_0,t)}{r_0}}\, \right]\times\d t.
\nonumber\\
\end{eqnarray}
That is:
\begin{eqnarray}
\hbox{(net flux)}_{t_\mathrm{initial}}^{t_\mathrm{final}} 
&=&\int\frac{\dot{m}(r_0,t)}{\sqrt{{1-\frac{2\,m(r_0,t)}{r}}}}\,\d t\nonumber\\
&=&\left[-r_0\sqrt{1-\frac{2\,m(r_0,t)}{r_0}} \,\right]_{t_\mathrm{initial}}^{t_\mathrm{final}}\!\!\!.
\end{eqnarray}
Then, if initially there is no mass inside $r=r_0$,  at time $t_\mathrm{final}$ we have
\begin{equation}
\hbox{(net flux)}_{t_\mathrm{initial}}^{t_\mathrm{final}} = r_0\,\left( 1-\sqrt{1-\frac{2\,m(r_0,t_\mathrm{final})}{r_0}}\,\right).
\end{equation}
In this situation, the only meaningful definition of internal energy at $t=t_\mathrm{initial}$ is to set $U(r_0,t_\mathrm{initial})=0$. Hence, at any subsequent time $t$ the internal energy $U(r_0,t)$ is equal to the net incoming flux and so it makes sense to define
\begin{equation}
\label{BYmass}
U(r_0,t)=r_0\,\left( 1-\sqrt{1-\frac{2\,m(r_0,t)}{r_0}}\,\right).
\end{equation}
This \emph{internal} energy is just the Brown--York quasi-local mass for the spacetime geometry with metric (\ref{met}), see  \cite{brown-york}. We can rearrange this (as pointed out in \cite{brown-york})  to yield
\begin{equation}
m(r_0,t)=U(r_0,t)-\frac{U^2(r_0,t)}{2\,r_0}.
\end{equation}
Here $m(r_0,t)$ retains its interpretation as the Hawking--Israel (and Hernandez--Misner/ Misner--Sharp~\cite{hernandez-misner, misner-sharp}) quasi-local mass. The difference between the two notions of energy is just the self interacting Newtonian gravitational potential of a massive shell of radius $r_0$. Both energies coincide at spatial infinity with the ADM mass. 

\section{Surface gravity}\label{S:surface-gravity}

Several attempts at  calculating the surface gravity for a time-dependent metric have been made using the Kodama vector instead of the Killing vector \cite{Hayward}, with results qualitatively similar to those in the static case; even to the extent of deriving some form of the first law of (black hole) thermodynamics. 

\subsection{Surface gravity from fiducial observers}

The most intuitive way of calculating the surface gravity is by working in the exterior region and considering the four-velocity $V$ parallel to the Kodama vector. Calculate the four-acceleration $A = \nabla_V V$. Then explicitly computing the magnitude of this four-acceleration we see
\begin{eqnarray}
\label{aa}
a=||A|| &=&  {1\over\sqrt{1-{2m(r,t)\over r}}} \left[ {m(r,t)\over r^2} - {m'(r,t)\over r} \right] 
 \qquad
\nonumber\\
&&  -\sqrt{1-{2m(r,t)\over r}}\,{\Phi'(r,t)}.
\end{eqnarray}
(Note that near spatial infinity we have the sensible Newtonian result $a\to m/r^2$.)
The surface gravity can be defined as the acceleration of an observer near the evolving horizon, which we implicitly define by $r_H(t)=2m(r_H(t),t)$, as measured by another observer at infinity. Thus, at this point it is necessary to multiply by a normalizing factor, often referred to as a redshift factor.  In the (asymptotically flat) static case there is no doubt that the appropriate normalizing factor is 
\begin{equation}
||K|| = |g_{tt}|^{1/2} = e^{-\Phi(r)} \sqrt{1-2m(r)/r}
\end{equation}
and that the appropriate object to consider is the near horizon limit of
\begin{eqnarray}
\kappa_\mathrm{static} &=& {||A||\;  ||K||} =  ||\nabla_K V|| 
\nonumber
\\
&=&
 e^{-\Phi(r)} \left\{ \left[ {m(r)\over r^2} - {m'(r)\over r} \right]  \right.
  \qquad
\nonumber\\
&&
\quad\quad\quad
\left. -\left[1-{2m(r)\over r}\right] \,{\Phi'(r)} \right\}.
\end{eqnarray}
The location of the horizon is in the static case implicitly defined by  $r_H=2m(r_H)$, and this is now a true Killing horizon and also an event horizon, at which we have the standard result~\cite{putative}
\begin{equation}
\label{E:static}
\kappa_{\mathrm{static}|H}  =  e^{-\Phi(r_H)} \left\{ {1-2m'(r_H)\over 2 r_H} \right\}.
\end{equation}
When bootstrapping to the dynamic case a plausible generalization (which we shall subsequently  buttress by also considering the radial null geodesics) is to replace $||K||\to||\T||$, which at least has the virtue of maintaining the correct static limit. Under this hypothesis the appropriate object to consider is the near horizon limit of
\begin{eqnarray}
\kappa_V &=& {||A||\;  ||\T||} =  ||\nabla_\T V|| 
\nonumber\\
&=&
  e^{-\Phi(r,t)} \left\{ \left[ {m(r,t)\over r^2} - {m'(r,t)\over r} \right] \right.
   \qquad
\nonumber\\
&&
\quad\quad\quad
\left.   -\left[1-{2m(r,t)\over r}\right] \,{\Phi'(r,t)} \right\}.
\end{eqnarray}
which on the evolving horizon has the limit
\begin{equation}
\kappa_{V|H}(t)  =  e^{-\Phi(r_H(t),t)} \left\{ {1- 2 m'(r_H(t),t)\over 2r_H(t)} \right\}.
\end{equation}
However, this proposed definition presents us with a potential  ambiguity --- what is the physically most appropriate choice for the normalizing factor? 
The choice of $||\T||$ as normalizing factor as advocated above is certainly plausible, and has the correct static limit. Furthermore it is intimately related to the Kodama time introduced in this article, rather than the Kodama vector $k$. Nevertheless, it is useful to see if we can come up with other plausible candidates for surface gravity in an evolving spacetime, and see whether they agree with (or are closely related to) the above proposal, and whether they possess the correct static limit.

\subsection{Surface gravity from radial null geodesics}

There are other, possibly less ambiguous, ways to usefully define the sought after surface gravity. More specifically, we can parameterize the strength of the  gravitational field throughout the entire spacetime geometry by using the inaffinity properties of the radial null geodesics. Consider (temporarily) the following null vectors:
\begin{equation}
\label{l1}
\tilde \ell^\pm_a={\pm k_a+\nabla_a r\over 2}.
\end{equation}
In the exterior region (where $k$ is timelike) these null vectors are both outward-pointing, $\tilde \ell^+$ is future-pointing, and $\tilde \ell^-$ is past-pointing. (Note that $-\tilde \ell^-$ is then inward pointing; these specific conventions have been chosen to simplify the computations below as far as possible.) 
These are arguably the simplest radial null vectors one could write down using only the Kodama vector. 
It is easy to check that $\tilde \ell^+_a \; \tilde \ell_-^a={1\over2} ||k||^2$. Since we are working with spherical symmetry, both radial null vectors must satisfy the geodesic equation (in its non-affine parameterized form):
\begin{equation}
\label{geo}
\tilde \ell_\pm^b\,\nabla_b\tilde \ell_\pm^a=\tilde \kappa_{\ell_\pm}\, \tilde\ell_\pm^a\,; 
\qquad\qquad 
\nabla_{\ell_\pm} \ell_\pm =  \tilde \kappa_{\ell_\pm}\, \tilde\ell_\pm\,; 
\end{equation}
where $\tilde \kappa_{\ell\pm}$ are scalars defined everywhere throughout the spacetime. By contracting these equations with $\tilde \ell^\mp_a$, we can explicitly compute $\tilde \kappa_{\ell_\pm}$, to yield:
\begin{equation}
\label{kappaL}
\tilde \kappa_{\ell_+}=\tilde \kappa_{\ell_-} ={m(r,t)\over r^2}-{m'(r,t)\over r}-{1\over2}\left[1-{2\,m(r,t)\over r}\right]\Phi'(r,t).
\end{equation}
(Note that near spatial infinity we have $\tilde\kappa_{\ell_\pm} \to m/r^2$.)
At the evolving horizon this would reduce to the tentative definition
\begin{equation}
\tilde \kappa_H(t)={1-2\,m'(r_H(t),t)\over 2 r_H(t)}.
\end{equation}
Unfortunately this does \emph{not} reduce to the known result in the static case --- there is a missing factor of $e^{-\Phi(r_H(t),t)}\to e^{-\Phi(r_H)}$.  This makes the above definition not suitable for calculating the Hawking temperature. (We emphasize that in static situations the standard Wick-rotation trick of going to Euclidean signature, demanding the absence of any conical singularity at $r_H$, and interpreting the Hawking temperature in terms of periodicity in imaginary time, uniquely enforces equation (\ref{E:static}) as the only physically acceptable candidate for the surface gravity~\cite{putative}.)  
The source of the difficulty is, since $e^{-\Phi(r,t)} = ||\T||/||k||$, ultimately due to the fact that $||\T|| \neq ||k||$ in general.

These considerations do suggest an improved strategy: Since we have seen how to use the Clebsch decomposition to deduce the natural existence of a Kodama time, in addition to a Kodama vector, then it would seem appropriate to use the Kodama time as the natural (non-affine) parameter for these radial null curves. (That is, we now parameterize the null curves by Kodama time, rather than the usual Killing time used in the static case.) 
This is tantamount to choosing
\begin{equation}
\label{nl2a}
\ell^\pm_a= {1\over2} \left[ \pm \T_a+  {||\T||\over||k||} \nabla_a r \right] = e^{-\Phi(r,t)} \;\tilde \ell^\pm_a;
\end{equation}
This time, the inner product is $ \ell^+_a \; \ell_-^a = {1\over2} ||\T||^2$. These ``Kodama time normalized'' radial null vectors are again tangent to the radial null geodesics and so satisfy 
\begin{equation}
\label{geo2}
 \ell^b_\pm\,\nabla_b\ell^a_\pm=\kappa_{\ell_\pm} \, \ell_\pm^a. 
\end{equation}
A brief calculation yields
\begin{equation}
\kappa_{\ell_\pm} = e^{-\Phi(r,t)} \; \tilde \kappa_{\ell_\pm} - \ell^a_\pm \nabla_a \Phi, 
\end{equation}
whence
\begin{eqnarray}
\label{bulk2}
{\kappa}_{\ell_\pm}&=&  e^{-\Phi(r,t)} \Bigg\{ \left[{m(r,t)\over r^2}-{m'(r,t)\over r}\right]
\nonumber\\
&&
\qquad 
-\left[1-{2\,m(r,t)\over r}\right]  {\Phi'(r)} \Bigg\} 
\mp {1\over2}\dot{\Phi}(r,t).
\quad
\end{eqnarray}
At  the evolving horizon, ${\kappa}_{\ell_\pm}$ reduces to
\begin{eqnarray}
\kappa_{\ell_\pm|H}(t)&=&  e^{-\Phi(r_H(t),t)} \left\{ {1-2\,m'(r_H(t),t)\over 2 r_H(t)} \right\} 
 \qquad
\nonumber\\
&&
\quad\quad\quad \quad
\mp {1\over2} \dot{\Phi}(r_H(t),t).
\end{eqnarray}
This is not quite equal to $\kappa_{V|H}$ --- though it does share with  $\kappa_{V|H}$ the desirable property of having the correct static limit. An improved proposal is to \emph{average} over past and future pointing null geodesics and take
\begin{equation}
\kappa_\mathrm{null} =  {1\over 2} \left[ \kappa_{\ell_+} + \kappa_{\ell_-} \right].
\end{equation}
Then $ \kappa_\mathrm{null}   = \kappa_V$. 
That is: If one takes future-pointing and past-pointing outward null geodesics, normalized to Kodama time, and averages the resulting inaffinity parameters, then one obtains the same $\kappa_V$ that we tentatively identified based on the 4-acceleration of the FIDOs that follow the Kodama flow. 

In short: By using Kodama time in addition to the Kodama vector we have now developed a geometrically preferred notion of surface gravity for spherically symmetric evolving spacetimes  that can meaningfully be extended throughout the entire spacetime, and in addition exhibits a good static limit.

\section{The evolving horizon}\label{S:evolving-horizon}

With the calculations presented so far, it is not possible to conclude too much about the evolving horizon at $r_H(t)=2\,m(t,r_H)$. To relate this to a trapping horizon, in the Hayward sense \cite{Hayward, Nielsen}, it is necessary to compute the expansions, $\theta_{\tilde\ell_\pm}$ and $\theta_{\ell_\pm}$, of the radial null vectors. Let us use the following definitions for the expansion \cite{Nielsen-Visser} 
\begin{equation}
\theta_{\tilde\ell_\pm}=\nabla_a\tilde \ell^a_\pm- \tilde \kappa_{\ell_\pm}; 
\qquad \qquad
\theta_{\ell_\pm}=\nabla_a \ell^a_\pm- \kappa_{\ell_\pm}; 
\end{equation} 
A brief computation yields
\begin{eqnarray}
\theta_{\tilde\ell_\pm}&=&{1\over r}\left(1-{2\,m(r,t)\over r}\right); 
\nonumber\\
 \theta_{\ell_\pm}&=&{1\over r}\left(1-{2\,m(r,t)\over r}\right)\; e^{-\Phi(r,t)}.
\end{eqnarray}
In particular both $\theta_{\tilde\ell_+}$ and $\theta_{{\ell_+}}$, change sign at the evolving horizon $r_H(t)$.  This is sufficient to guarantee that the evolving horizon at $r_H(t)=2\,m(t,r_H(t))$ is an apparent horizon. However $r_H(t)$ is not a trapping horizon in the Hayward sense~\cite{Hayward}, nor an ``Ashtekar horizon"~\cite{Ashtekar}, since in addition $\theta_{\tilde{\ell}_{-}}$ and $\theta_{\ell_{-}}$ also both vanish on the evolving horizon. This is not critical for our purposes, since ultimately a trapping horizon is not needed to have Hawking radiation \cite{essential, no-trap, quasiparticle}.

\smallskip

For completeness, we have also computed $\tilde{\ell}^a_{-}\nabla_a\theta_{\tilde{\ell}_{+}}$ and $\ell^a_{-}\nabla_a\theta_{\ell_{+}}$. 
At the evolving horizon we have
\begin{eqnarray}
\left(\tilde \ell_-^a\nabla_a\,\theta_{\tilde\ell_+}\right)_{H}&=& {\dot{m}(r_H(t),t)\over r_H(t)^2}\,e^{\Phi(r_H(t),t)};
\nonumber\\
\left(\ell_-^a\nabla_a\,\theta_{\ell_+}\right)_{H}&=&{\dot{m}(r_H(t),t)\over r_H(t)^2}\,e^{-\Phi(r_H(t),t)}.
\end{eqnarray}
Thus for both normalizations we have most (but not all) of the key properties of an outer trapping horizon at $r_H(t)$ when $\dot{m}(t,r_H(t))>0$, \emph{i.e.}, when the overall mass increases in time. 

\section{Discussion}\label{S:discussion}
We have used the warped product formalism to investigate the geometry of time-dependent spherically symmetric spacetimes, developing relatively straightforward arguments for the covariant conservation of the Kodama vector itself and the associated Kodama flux. This construction has allowed us to construct a very general class of conserved fluxes appropriate to any spherically symmetric spacetime. 

Furthermore, we have successfully used the Kodama vector field, plus the Clebsch decomposition,  to obtain a preferred Kodama time coordinate, and have then proceeded to construct a  geometrically preferred coordinate system for describing spherically symmetric time dependent spacetimes. The resulting metric is one of the most simple forms of the metric of a spherically symmetric spacetime --- a diagonal metric in Schwarzchild curvature coordinates. With these coordinates there are very simple physical interpretations for both the Hawking--Israel (Hernandez--Misner/ Misner--Sharp) and Brown--York quasi-local masses. Although the definition of surface gravity remains somewhat ambiguous, by using the Kodama time as an integral part of the construction we have identified some very good geometrically preferred candidates that are compatible with known results in the static limit.



\end{document}